\documentstyle[twoside,epsfig]{article}

\catcode`\@=11
\long\def\@makefntext#1{
\protect\noindent \hbox to 3.2pt {\hskip-.9pt  
$^{{\eightrm\@thefnmark}}$\hfil}#1\hfill}		

\def\@makefnmark{\hbox to 0pt{$^{\@thefnmark}$\hss}}	
	
\def\ps@myheadings{\let\@mkboth\@gobbletwo
\def\@oddhead{\hbox{}
\rightmark\hfil\eightrm\thepage}   
\def\@oddfoot{}\def\@evenhead{\eightrm\thepage\hfil
\leftmark\hbox{}}\def\@evenfoot{}
\def\sectionmark##1{}\def\subsectionmark##1{}}

\oddsidemargin=\evensidemargin
\newcommand{\be}{\begin{equation}}    
\newcommand{\ee}{\end{equation}}
\newcommand{\beq}{\begin{eqnarray}}
\newcommand{\eeq}{\end{eqnarray}}
\newcommand{\dps}{\displaystyle}

\def\op{ \ $ }
\def\cl{$ \ }

\def\msun{M_\odot}
\def\nn{\nonumber}

\newcounter{sectionc}\newcounter{subsectionc}\newcounter{subsubsectionc}
\renewcommand{\section}[1] {\vspace{12pt}\addtocounter{sectionc}{1} 
\setcounter{subsectionc}{0}\setcounter{subsubsectionc}{0}\noindent 
	{\tenbf\thesectionc. #1}\par\vspace{5pt}}
\renewcommand{\subsection}[1] {\vspace{12pt}\addtocounter{subsectionc}{1} 
	\setcounter{subsubsectionc}{0}\noindent 
	{\bf\thesectionc.\thesubsectionc. {\kern1pt \bfit #1}}\par\vspace{5pt}}
\renewcommand{\subsubsection}[1] {\vspace{12pt}\addtocounter{subsubsectionc}{1}
	\noindent{\tenrm\thesectionc.\thesubsectionc.\thesubsubsectionc.
	{\kern1pt \tenit #1}}\par\vspace{5pt}}
\newcommand{\nonumsection}[1] {\vspace{12pt}\noindent{\tenbf #1}
	\par\vspace{5pt}}

\newcounter{appendixc}
\newcounter{subappendixc}[appendixc]
\newcounter{subsubappendixc}[subappendixc]
\renewcommand{\thesubappendixc}{\Alph{appendixc}.\arabic{subappendixc}}
\renewcommand{\thesubsubappendixc}
	{\Alph{appendixc}.\arabic{subappendixc}.\arabic{subsubappendixc}}

\renewcommand{\appendix}[1] {\vspace{12pt}
        \refstepcounter{appendixc}
        \setcounter{figure}{0}
        \setcounter{table}{0}
        \setcounter{lemma}{0}
        \setcounter{theorem}{0}
        \setcounter{corollary}{0}
        \setcounter{definition}{0}
        \setcounter{equation}{0}
        \renewcommand{\thefigure}{\Alph{appendixc}.\arabic{figure}}
        \renewcommand{\thetable}{\Alph{appendixc}.\arabic{table}}
        \renewcommand{\theappendixc}{\Alph{appendixc}}
        \renewcommand{\thelemma}{\Alph{appendixc}.\arabic{lemma}}
        \renewcommand{\thetheorem}{\Alph{appendixc}.\arabic{theorem}}
        \renewcommand{\thedefinition}{\Alph{appendixc}.\arabic{definition}}
        \renewcommand{\thecorollary}{\Alph{appendixc}.\arabic{corollary}}
        \renewcommand{\theequation}{\Alph{appendixc}.\arabic{equation}}
        \noindent{\tenbf Appendix \theappendixc #1}\par\vspace{5pt}}
\newcommand{\subappendix}[1] {\vspace{12pt}
        \refstepcounter{subappendixc}
        \noindent{\bf Appendix \thesubappendixc. {\kern1pt \bfit #1}}
	\par\vspace{5pt}}
\newcommand{\subsubappendix}[1] {\vspace{12pt}
        \refstepcounter{subsubappendixc}
        \noindent{\rm Appendix \thesubsubappendixc. {\kern1pt \tenit #1}}
	\par\vspace{5pt}}

\newcommand{\textlineskip}{\baselineskip=13pt}
\newcommand{\smalllineskip}{\baselineskip=10pt}

\def\eightcirc{
\begin{picture}(0,0)
\put(4.4,1.8){\circle{6.5}}
\end{picture}}
\def\eightcopyright{\eightcirc\kern2.7pt\hbox{\eightrm c}} 

\newcommand{\copyrightheading}[1]
	{\vspace*{-2.5cm}\smalllineskip{\flushleft
	{\footnotesize International Journal of Modern Physics D, #1}\\
	{\footnotesize $\eightcopyright$\, World Scientific Publishing
	 Company}\\
	 }}

\newcommand{\publisher}[2]{{\begin{center}\footnotesize\smalllineskip 
	Received #1\\
	Revised #2
	\end{center}
	}}

\def\abstracts#1#2#3{{
	\centering{\begin{minipage}{4.5in}\baselineskip=10pt\footnotesize
	\parindent=0pt #1\par 
	\parindent=15pt #2\par
	\parindent=15pt #3
	\end{minipage}}\par}} 

\renewenvironment{thebibliography}[1]
        {\frenchspacing
	 \ninerm\baselineskip=11pt
         \begin{list}{\arabic{enumi}.}
        {\usecounter{enumi}\setlength{\parsep}{0pt}     
	 \setlength{\leftmargin 12.7pt}{\rightmargin 0pt} 
         \setlength{\itemsep}{0pt} \settowidth
	{\labelwidth}{#1.}\sloppy}}{\end{list}}

\newcounter{itemlistc}
\newcounter{romanlistc}
\newcounter{alphlistc}
\newcounter{arabiclistc}

\newcommand{\fcaption}[1]{
        \refstepcounter{figure}
        \setbox\@tempboxa = \hbox{\footnotesize Fig.~\thefigure. #1}
        \ifdim \wd\@tempboxa > 5in
           {\begin{center}
        \parbox{5in}{\footnotesize\smalllineskip Fig.~\thefigure. #1}
            \end{center}}
        \else
             {\begin{center}
             {\footnotesize Fig.~\thefigure. #1}
              \end{center}}
        \fi}

\newcommand{\tcaption}[1]{
        \refstepcounter{table}
        \setbox\@tempboxa = \hbox{\footnotesize Table~\thetable. #1}
        \ifdim \wd\@tempboxa > 5in
           {\begin{center}
        \parbox{5in}{\footnotesize\smalllineskip Table~\thetable. #1}
            \end{center}}
        \else
             {\begin{center}
             {\footnotesize Table~\thetable. #1}
              \end{center}}
        \fi}

\def\@citex[#1]#2{\if@filesw\immediate\write\@auxout
	{\string\citation{#2}}\fi
\def\@citea{}\@cite{\@for\@citeb:=#2\do
	{\@citea\def\@citea{,}\@ifundefined
	{b@\@citeb}{{\bf ?}\@warning
	{Citation `\@citeb' on page \thepage \space undefined}}
	{\csname b@\@citeb\endcsname}}}{#1}}

\newif\if@cghi
\def\cite{\@cghitrue\@ifnextchar [{\@tempswatrue
	\@citex}{\@tempswafalse\@citex[]}}
\def\citelow{\@cghifalse\@ifnextchar [{\@tempswatrue
	\@citex}{\@tempswafalse\@citex[]}}
\def\@cite#1#2{{$\null^{#1}$\if@tempswa\typeout
	{IJCGA warning: optional citation argument 
	ignored: `#2'} \fi}}

\def\pmb#1{\setbox0=\hbox{#1}
	\kern-.025em\copy0\kern-\wd0
	\kern.05em\copy0\kern-\wd0
	\kern-.025em\raise.0433em\box0}

\def\fnt#1#2{\footnotetext{\kern-.3em
	{$^{\mbox{\scriptsize #1}}$}{#2}}}

\def\fpage#1{\begingroup
\voffset=.3in
\thispagestyle{empty}\begin{table}[b]\centerline{\footnotesize #1}
	\end{table}\endgroup}

\def\runninghead#1#2{\pagestyle{myheadings}
\markboth{{\protect\footnotesize\it{\quad #1}}\hfill}
{\hfill{\protect\footnotesize\it{#2\quad}}}}
\font\tenrm=cmr10
\font\tenit=cmti10 
\font\tenbf=cmbx10
\font\bfit=cmbxti10 at 10pt
\font\ninerm=cmr9

\font\eightrm=cmr8

\def\qed{\hbox{${\vcenter{\vbox{			
   \hrule height 0.4pt\hbox{\vrule width 0.4pt height 6pt
   \kern5pt\vrule width 0.4pt}\hrule height 0.4pt}}}$}}

\begin{document}

\runninghead{Gravitational waves emitted by extrasolar
planetary systems $\ldots$}{Gravitational waves emitted by extrasolar
planetary systems $\ldots$} 

\normalsize\textlineskip
\thispagestyle{empty}
\setcounter{page}{1}

\copyrightheading{}			

\vspace*{0.88truein}

\fpage{1}
\centerline{\bf GRAVITATIONAL WAVES EMITTED BY  EXTRASOLAR}
\vspace*{0.035truein}
\centerline{\bf PLANETARY SYSTEMS}
\vspace*{0.37truein}
\centerline{\footnotesize VALERIA FERRARI}
\vspace*{0.015truein}
\centerline{\footnotesize\it 
Dipartimento di Fisica ``G. Marconi",
Universit\'a degli Studi di Roma, ``La Sapienza"}
\centerline{\footnotesize\it and} 
\centerline{\footnotesize\it Sezione INFN  ROMA1, piazzale Aldo  Moro 2}
\baselineskip=10pt
\centerline{\footnotesize\it 00185 Roma, Italy}

\vspace*{10pt}
\centerline{\footnotesize MARCO D'ANDREA}
\vspace*{0.015truein}
\centerline{\footnotesize\it 
Dipartimento di Fisica ``G. Marconi",
Universit\'a degli Studi di Roma, ``La Sapienza"
}
\baselineskip=10pt
\centerline{\footnotesize\it 00185 Roma, Italy}

\vspace*{10pt}
\centerline{\footnotesize EMANUELE BERTI}
\vspace*{0.015truein}
\centerline{\footnotesize\it 
Dipartimento di Fisica ``G. Marconi",
Universit\'a degli Studi di Roma, ``La Sapienza"}
\centerline{\footnotesize\it and} 
\centerline{\footnotesize\it Sezione INFN  ROMA1, piazzale Aldo  Moro 2}
\baselineskip=10pt
\centerline{\footnotesize\it 00185 Roma, Italy}

\vspace*{0.225truein}
\publisher{(received date)}{(revised date)}

\vspace*{0.21truein}
\abstracts{
In this paper we consider the Extra-solar Planetary Systems 
recently discovered in our Galaxy as potential sources of gravitational
waves.
We  estimate  the frequency and characteristic amplitude of
the radiation they emit  due to the orbital 
motion, using the quadrupole formalism.
In addition, we check whether the conditions needed for the resonant 
excitation of the $f$- and $g$-modes of the central star can be 
fulfilled.  By a Roche-lobe analysis, we show that there could exist
systems in which the low-order $g$-modes could be excited,
although this does not happen  in the systems discovered up to now. 
}{}{}



\vspace*{1pt}\textlineskip	
\section{Introduction}	
\vspace*{-0.5pt}
\noindent

Since 1992, when the first Extra-solar Planetary System (EPS)
was discovered\cite{Wolszczan},
a number of such  systems, composed of one or more planets
orbiting around a main sequence solar-type star
has been observed\cite{Uff2}, 
and it is conceivable that many others will
be found in the near future.  
Some of the discovered EPS's are well characterized, since the 
mass and radius of the main star,   the mass of the
observed planets and  their orbital parameters can be deduced from
observations.
With this information it is possible to attempt  
a first estimate of the characteristics of
the gravitational radiation emitted by these systems, that, being at a 
distance of a few tens of parsecs, are  very close to us. 
We shall select a set of EPS's for which all needed information is available, 
and compute the gravitational signal emitted by each couple
star-planet due to the orbital motion, 
using the quadrupole formalism. 
The estimated frequency and amplitude of the waves 
impinging on Earth will be compared 
with the emission of the binary pulsar 
PSR 1913+16\cite{hulsetaylor}$^,$\cite{TaylorWeisberg}$^,$\cite{thorsett}.

We shall then consider a  further mechanism of gravitational emission which
may play an interesting role in these systems, i.e. the resonant
excitation of the modes of oscillation of the star induced by tidal
interaction with the orbiting planet. This problem has been studied 
in the literature for solar
type stars with a planet, and for neutron star-neutron star close 
binary systems by using different approaches, based essentially 
on the  deformation induced by the dynamical tides raised  by the 
companion\cite{alexander}$^,$\cite{terquem}$^,$\cite{shaferkostas}.
The emission of gravitational waves associated with the resonant excitation 
of the modes of a star has been studied   by Kojima\cite{kojima}.
He considered a small  mass in circular orbit
around a compact star, and integrated the equations  describing the
perturbation of the star excited by the small mass, in general relativity. 
He showed that a sharp resonance occurs if the frequency 
of the fundamental mode of the star is twice the orbital  frequency,
and that the characteristic wave amplitude emitted at that frequency
can be up to
100 times larger than that evaluated by the quadrupole formula.
In this view, it is interesting to check whether
the conditions of resonant  excitation
can be fulfilled in planetary systems for some modes of the central star, 
and in particular if
this is the case for any of the EPS's listed in table 1.
This will be done in section 4.
In addition, we shall discuss the  possibility for a planet
to get sufficiently close to a star to excite the lowest order
$g$-modes, and possibly the fundamental one,
without being disrupted by tidal forces. We will show
that for solar type stars the resonant excitation of the low-order
$g$-modes may, in principle, be possible.

\section{Main characteristics of the extrasolar planetary systems}

Among the EPS's discovered up to now,  we have selected  those
for which the  parameters of the central star and of the planets,
which we need to estimate the gravitational emission, have
been determined with sufficient accuracy.
These parameters are tabulated in table 1 and 2.
Here and in the following,
data  will  be given  with the corresponding errors, 
when available in the literature.
In the first column of  table 1 we list the 
selected EPS's with the corresponding bibliography, and in column 2 the
spectral class of the central stars.
Most of them  belong to  a class
similar to that of the Sun, which is a G2 V star.

\begin{table}
\centering
\caption{The spectral class of the  central star,
its mass, $M_\star$, its distance from Earth, $D$,
the factor $\sqrt{\frac{GM_\star}{R_\star^3}}$ (rad$\cdot$s$^{-1}$),
and the ratio between the
lower limit of the estimated mass of the planet, $M_{p}\sin{i}$,
and $M_\star$, are tabulated for a selected set of EPS's.
Strictly speaking, the bodies orbiting the stars listed below HD 114762 are
classified as brown dwarfs rather than as planets.}
\begin{tabular}{@{}llllcl@{}}
\\
\multicolumn{1}{c}{Star}&Spectral Cl. &$M_* (M_\odot)$  &  D (pc) 
& $\sqrt{\frac{GM_\star}{R_\star^3}}\cdot 10^4$ & $M_{p}\sin{i}/M_{\star}$\\
\hline

HD $75289$\protect\cite{UniGe}
&G$0$ V     & $1.05$    &$28.94$   &$5.4$    &$3.8\cdot10^{-4}$   \\

$51$ Peg\protect\cite{Ford}
&G$5$ V   &$1.05\pm0.09$   &$15.36$  &$5.1\pm0.7$       &$4.2\cdot10^{-4}$ \\

$\upsilon$ And\protect\cite{Ford}   
&F$8$ V  &$1.34\pm0.12$  &$13.47$   &$3.7\pm0.5$        &$5.3\cdot10^{-4}$ \\
         &           &               &   &   &$1.5\cdot10^{-3}$ \\
         &           &               &   &   &$3.3\cdot10^{-3}$ \\

$55$ Cnc\protect\cite{Ford}         
&G$8$ V     &$0.95\pm0.10$     &$12.53$  &$6.6\pm1.0$     &$8.8\cdot10^{-4}$ \\

$\rho$ CrB\protect\cite{Ford}       
&G$2$ V  &$0.89\pm0.05$  &$17.43$  &$3.8\pm0.5$  &$1.1\cdot10^{-3}$ \\

HD $210277$\protect\cite{MBV}
&G$0$ V    &$0.92$ &$21.29$  &$5.0$     &$1.3\cdot10^{-3}$  \\

$16$ Cyg B\protect\cite{Ford}       
&G$5$ V  &$0.96\pm0.05$ &$21.62$  &$5.0\pm0.6$      &$2.0\cdot10^{-3}$\\

Gl $876$\protect\cite{Uff2}           
&M$4$ V   &$ 0.3$  &$4.70$   &$13$          &$ 6.7\cdot10^{-3}$ \\

$47$ Uma\protect\cite{Ford}         
&G$0$ V   &$1.01\pm0.05$ &$14.08$ &$4.5\pm0.5$     &$2.2\cdot10^{-3}$  \\

$14$ Her\protect\cite{MBV}         
&K$0$ V     &$0.8$  &$18.15$   &$8.0$   &$3.9\cdot10^{-3}$ \\

Gl 86\protect\cite{MBV}            
&K1 V   &0.79      &10.91      &$8.0$        &$4.4\cdot10^{-3}$\\

$\tau$ Boo\protect\cite{Ford}       
&F$7$ V   &$1.37\pm0.09$ &$15.60$  &$4.7\pm0.6$    &$3.3\cdot10^{-3}$  \\

HD $168443$\protect\cite{MBV}        
&G$5$ V   &$0.94$   &$37.88$    &$7.2$     &$5.1\cdot10^{-3}$\\

$70$ Vir\protect\cite{Ford}         
&G$5$ V   &$1.01\pm0.05$  &$18.11$  &$2.5\pm0.2$    &$6.9\cdot10^{-3}$ \\

HD $114762$\protect\cite{Ford}      
&F$9$ V    &$0.75\pm0.15$ &$40.57$  &$4.0\pm1.0$    &$1.3\cdot10^{-2}$ \\

HD $110833$\protect\cite{MQU}
&K$3$ V    &$0.75$   &$17$  &$8.2$     &$2.2\cdot10^{-2}$ \\

HD $112758$\protect\cite{MQU}        
&K$0$ V     &$0.8$   &$16.5$  &$8.0$   &$4.2\cdot10^{-2}$  \\

HD $29587$\protect\cite{MQU}         
&G$2$ V     &$1.0$   &$45$   &$6.1$   &$3.8\cdot10^{-2}$\\

HD $283750$\protect\cite{Uff2}        
&K$2$ V     &$0.75$  &$16.5$  &$8.4$   &$6.4\cdot10^{-2}$  \\

HD $89707$\protect\cite{DM}         
&G$1$ V     &$1.2$  &$25$   &$6.1$    &$5.0\cdot10^{-2}$ \\

HD $217580$\protect\cite{MQU}        
&K$4$ V     &$0.7$  &$18$   &$9.1$    &$8.2\cdot10^{-2}$  \\
\hline
\end{tabular}
\end{table}
In column 3, 4 and 5 we tabulate, respectively, 
the mass of the central star, $M_\star,$  its distance from Earth, $D$,
- taken from the {\it Extrasolar Planets Catalog}\cite{Uff2}-
and the quantity $\sqrt{\frac{GM_\star}{R_\star^3}}$,
which will be used in section 4. 
It should be mentioned that 
the presence of planets  has been discovered mainly by
using accelerometric techniques, that is to say, by  measuring the
variations of the radial velocity of the star caused by its gravitational
interaction with the planet.
These techniques do not allow to determine the mass of the planet, $M_p$, 
but only the product $M_p \sin i$, where \op i \cl is the  angle between
the line of sight and the normal to the orbital plane. 
In column 6  we list the values $M_p \sin i$ for each planet, 
normalized to the mass of the central star.
From the data of table 1 we see that 
the closest system is at $4.70$ pc, the farthest at $45$ pc,
whereas the mass of the central star ranges within \op [0.3,1.37]~\msun.\cl

In table 2 we tabulate the orbital parameters of the planets
orbiting around the stars listed in table 1, i.e. the orbital
period $P$, the semimajor axis $a$,  and the eccentricity $e$.
From the data in column 2 we see that in
some cases the  period is very short, 
indicating that the planet gets very
close to the central star. For instance, it can be as short as 1.79 days
for the planet orbiting around HD 283750. 

\begin{table}
\centering
\caption{Orbital parameters of the planets orbiting around the stars listed in
table 1 (1 AU = $1.495978706\cdot10^{13}~\mbox{cm}$).}
\begin{tabular}{@{}lllll@{}}
\\
\multicolumn{1}{c}{Star}   &$P$     &$a~(\mbox{AU})$     &$e$  \\

\hline
 
HD $75289$      
&$3.5103\pm0.023\;d$   &0.046                    &$0.054\pm0.026$ \\

$51$ Peg       
&$4.2293\pm0.0011\;d$  &$0.0520\pm0.0015$        &$0.012\pm0.01$    \\

$\upsilon$ And   
&$4.6170\pm0.0003\;d$  &$0.598\pm0.018$           &$0.109\pm0.04$ \\
&$241.2\pm1.1\;d$      &0.83                      &$0.18\pm0.11$      \\
&$1266.6\pm30\;d$      &2.50                      &$0.41\pm0.11$    \\

$55$ Cnc                                        
&$14.648\pm0.0009\;d$  &$0.1152\pm0.0038$           &$0.051\pm0.13$     \\

$\rho$ CrB                                      
&$39.645\pm0.088\;d$   &$0.2192\pm0.0042$            &$0.15\pm0.03$  \\

HD $210277$                                     
&$437\pm25\;d$         &1.097                     &$0.45\pm0.08$       \\

$16$ Cyg B                                      
&$804\pm11.7\;d$       &$1.66\pm0.05$             &$0.634\pm0.082$   \\

Gl $876$                                        
&$60.85\pm0.15\;d$     &$0.21\pm0.01$             &$0.27\pm0.03$ \\

$47$ Uma                                        
&$3.0\;y$              &$2.08\pm0.06$             &$0.03\pm0.06$   \\

$14$ Her                                        
&$1619\pm70\;d$        &2.5                       &$0.3537\pm0.088$    \\

Gl $86$                                         
&$15.83\;d$            &0.11                      &$0.05$   \\

$\tau$ Boo                                      
&$3.3128\pm0.0002\;d$  &$0.0413\pm0.001$         &$0.018\pm0.016$ \\

HD $168443$                                     
&$57.9\;d$             &$0.277$                   &$0.54$         \\

$70$ Vir                                        
&$116.6\pm0.01\;d$     &$0.47\pm0.01$             &$0.4\pm0.01$\\

HD $114762$                                     
&$85.03\pm0.08;d$      &$0.34\pm0.04$             &$0.35\pm0.05$   \\

HD $110833$                                     
&$270.04\;d$           &$0.8$                 &$0.69$     \\

HD $112758$                                     
&$103.22\;d$           &$0.35$                &$0.16$          \\

HD $29587$                                      
&$3.17\;y$             &$2.5$                 &0   \\

HD $283750$                                     
&$1.79\;d$             &$0.04$                &$0.02$    \\

HD $89707$                                      
&$298.48\;d$           &$0.91$                &$0.93$   \\

HD $217580$                                     
&$456.44\;d$           &$1$                   &$0.52$ \\
\hline
 
\end{tabular}
\end{table}

\section{Gravitational wave emission due to the orbital motion}

We shall first evaluate the amount of gravitational radiation 
that each couple star-planet emits because of the orbital motion. 
Since we are  not interested in the
detailed form of the signal, the quadrupole formalism is sufficient
to derive the required information.
It is known that two pointlike masses revolving around their 
common center of mass  emit  gravitational
energy because of their time varying quadrupole moment.
To describe their motion it is convenient
to choose a coordinate system  with the
origin   located  at the center of mass of the system,
the orbital plane coincident with  the $x$-$y$ plane 
and  the $x$-axis oriented along the relative position vector,
\mbox{${\mathbf{X}}\equiv {\mathbf{x}}_{\star}-{\mathbf{x}}_{p}$},
when the planet is at the periastron.
The vector ${\mathbf{X}}$  describes, in general,
an ellipse with semi-major axis $a$ and eccentricity $e$, 
and its evolution is described by the parametric equations:
\begin{eqnarray}
\rho & = & a(1-e\cos{u}) \label{eq:radmo}\\
\alpha & = & 2\arctan{\left[\left(\frac{1+e}{1-e}\right)^{1/2}
\tan\frac{u}{2}\right]}. \label{eq:angmo}
\end{eqnarray}
Here $\rho = \left|{\mathbf{X}}\right|$,  $\alpha$ is the angle 
between ${\mathbf{X}}$ and the $x-$axis, and $u$ is the eccentric 
anomaly, related to time by the Kepler equation:
\be
\omega_k t=u-e\sin{u},
\label{inversion}
\ee
where \op \omega_k\cl is the keplerian orbital frequency
\[
\omega_k=\frac{2\pi}{P}=\left(\frac{G M}{a^3}\right)^{1/2},\]
and \op M=M_\star+M_p\cl is the total mass of the system.
From the reduced quadrupole moment $Q_{kl}$, given by:
\be
Q_{kl}= \mu\left(X^{k}X^{l}-\frac{1}{3}\delta^{k}_{l}
\left|{\mathbf{X}}\right|^{2}\right)
\ee
where \op\mu=M_\star M_p/M\cl is the reduced mass,
it is straightforward to compute the amplitudes 
of the metric perturbation, projected on a sphere at distance $r$ 
from the source:
\be
h_{ij}(t,r,\theta,\phi)=
\left.\frac{2G}{c^4 r}\left(P_{ik}P_{jl}-\frac{1}{2}P_{ij}P_{kl}\right)
\ddot{Q}_{kl}(\tau)\right|_{\tau=t-r/c}
\ee
where $P_{ij}=\delta^i_j-n_i n_j$ is the projector onto the plane 
orthogonal to the radial unit vector ${\bf n}$. 
The explicit expressions of \op h_{\theta \theta}\cl and \op h_{\theta \phi}\cl evaluated by this procedure are
\beq
h_{\theta \theta}(u,r,\theta,\phi)&=&\frac{2G^2M\mu}{c^4 ra(1-e\cos u)^3}
\times \\
\nn
&\times& \{(e^2-1)(1+\cos^2\theta)(1-2\cos^2\phi) \\
\nn
&+&\cos^2 u(e\cos u-2)\left[
(1-e^2)(\cos^2\phi+\cos^2\theta\cos^2\phi-\cos^2\theta)\right. \\
\nn
&+&\left.\cos^2\phi+\cos^2\theta\cos^2\phi-1 \right] 
+e\cos u(\cos^2\phi+\cos^2\theta\cos^2\phi-1) \\
\nn
&+&2\sqrt{1-e^2}\sin\phi\cos\phi(1+\cos^2\theta)
\sin u(e \cos^2 u-2\cos u+e)\},
\eeq

\beq
h_{\theta \phi}(u,r,\theta,\phi)&=&
\frac{4G^2M\mu\cos\theta}{c^4 ra(1-e\cos u)^3}
\times \\
\nn
&\times&\{\sin\phi\cos\phi\left[(e^2-2)
\cos^2 u(e\cos u-2)-e\cos u+2(e^2-1)\right] \\
\nn
&+&(2\cos^2\phi-1)\sqrt{1-e^2}\sin u (e\cos^2 u-2\cos u+e)\}.
\eeq
If the orbit is circular 
the radiation is  emitted at twice the orbital frequency
\op(\omega^{GW}=2\omega_k).\cl
By Fourier-transforming the wave amplitudes, and by averaging over the
solid angle, it is easy to  show that if the orbit is eccentric, waves
will be emitted at frequencies multiple of \op
\omega_k,\cl and the number of  equally spaced spectral lines will increase
with the eccentricity. 
\begin{figure}
\centerline{\mbox{
\psfig{figure=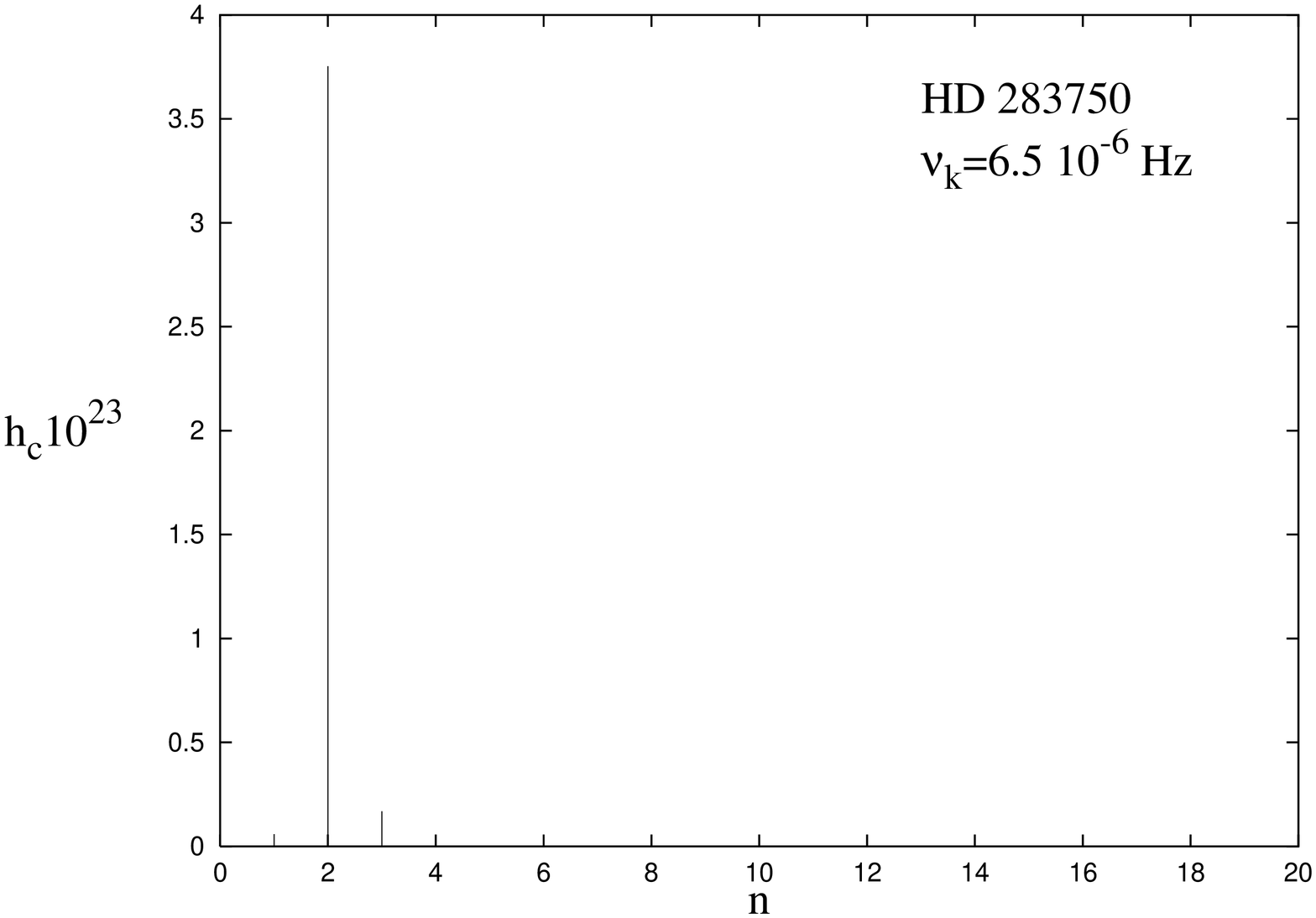,angle=270,width=9cm}~~~~~
\psfig{figure=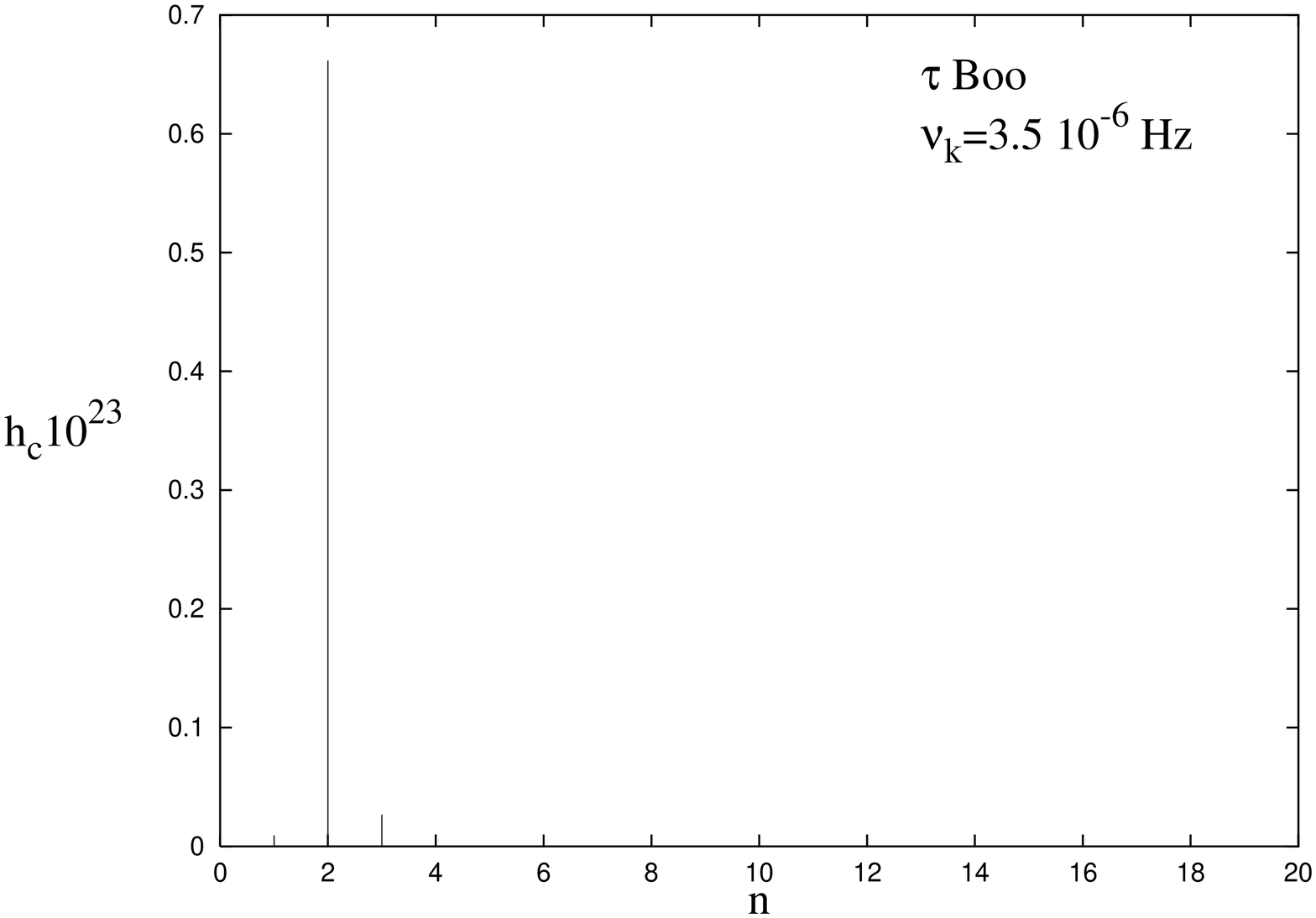,angle=270,width=9cm}}}
\vskip 8pt
\centerline{\mbox{      
\psfig{figure=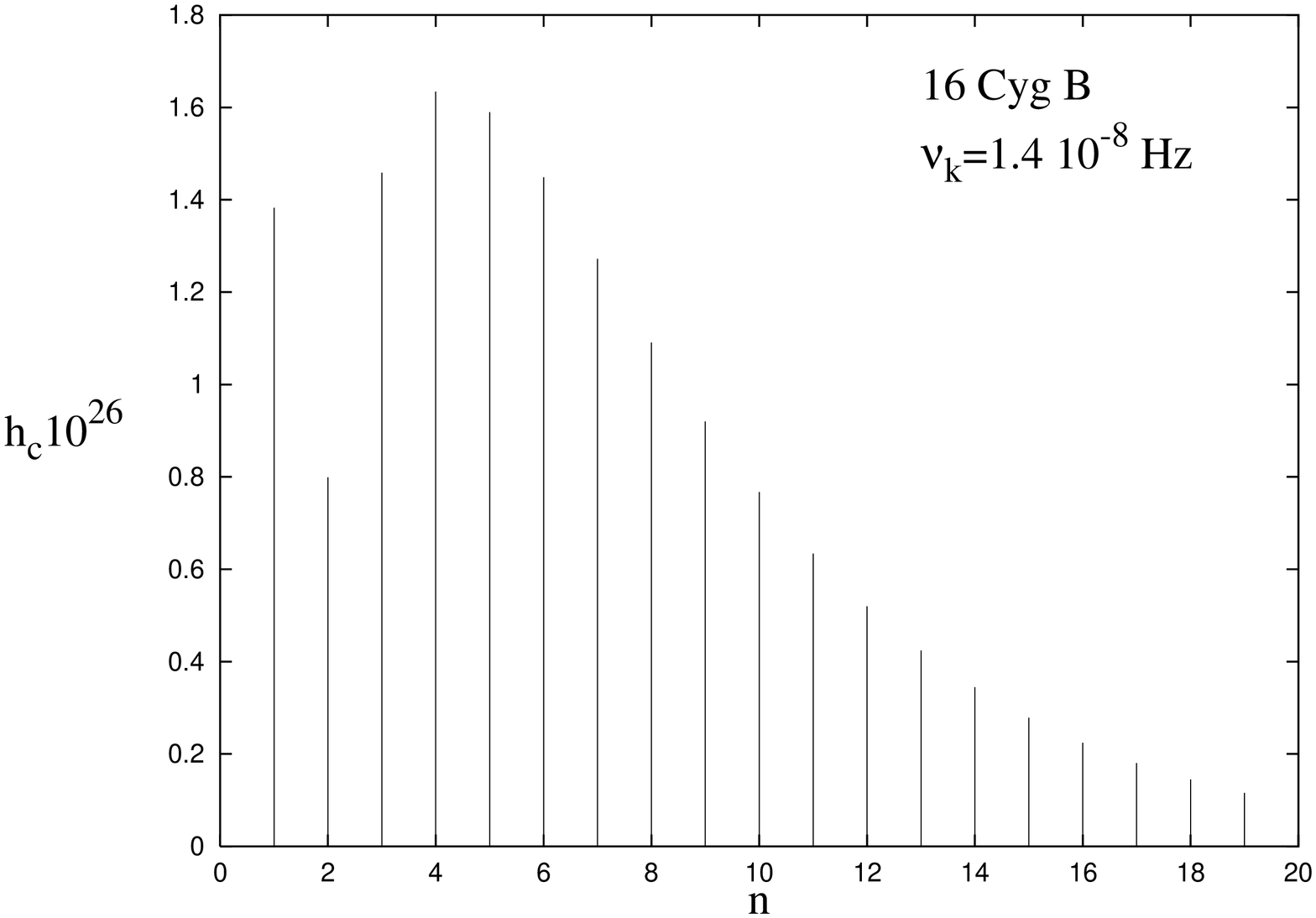,angle=270,width=9cm}~~~~~
\psfig{figure=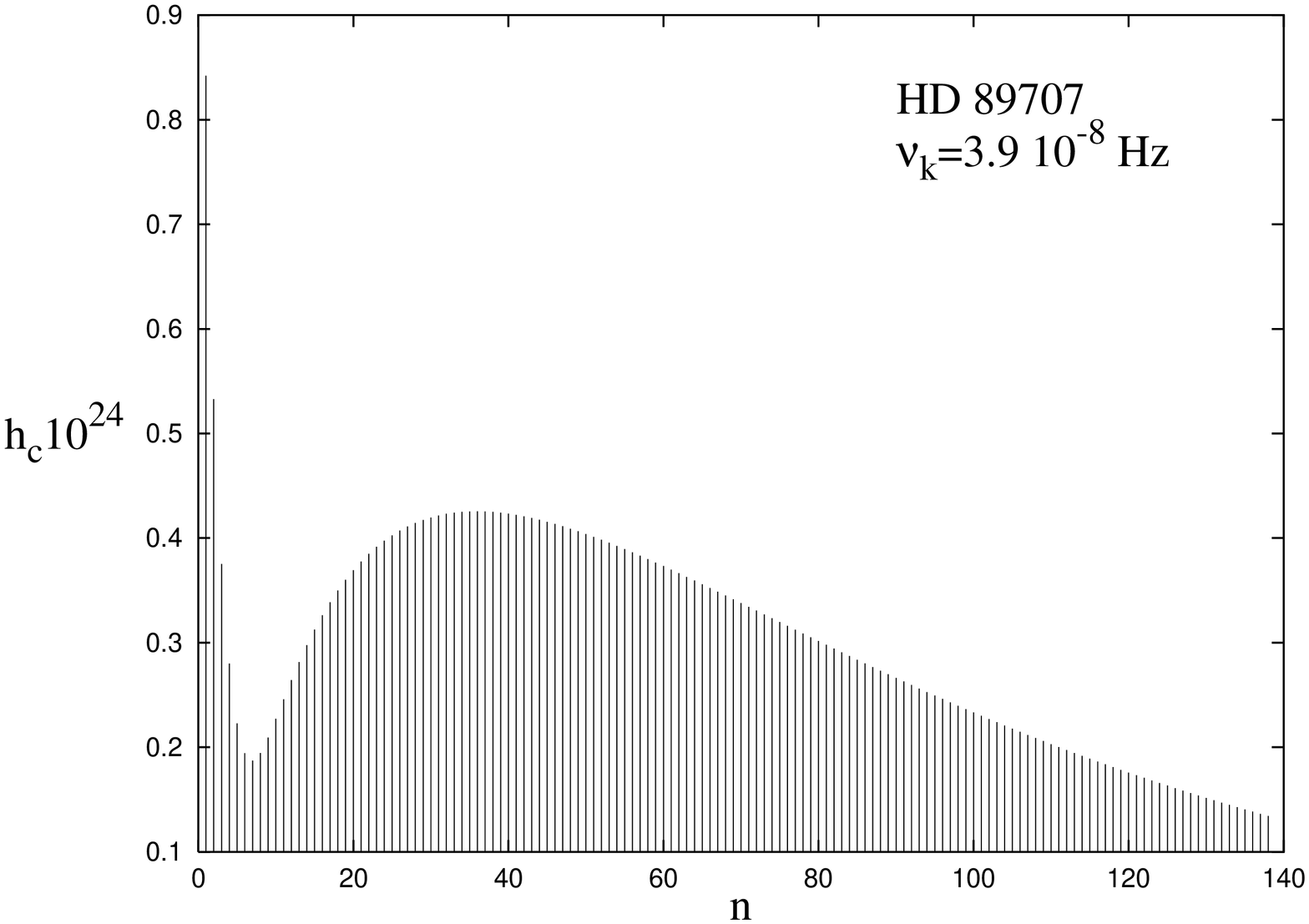,angle=270,width=9cm}}}
\vskip 8pt
\centerline{\mbox{
\psfig{figure=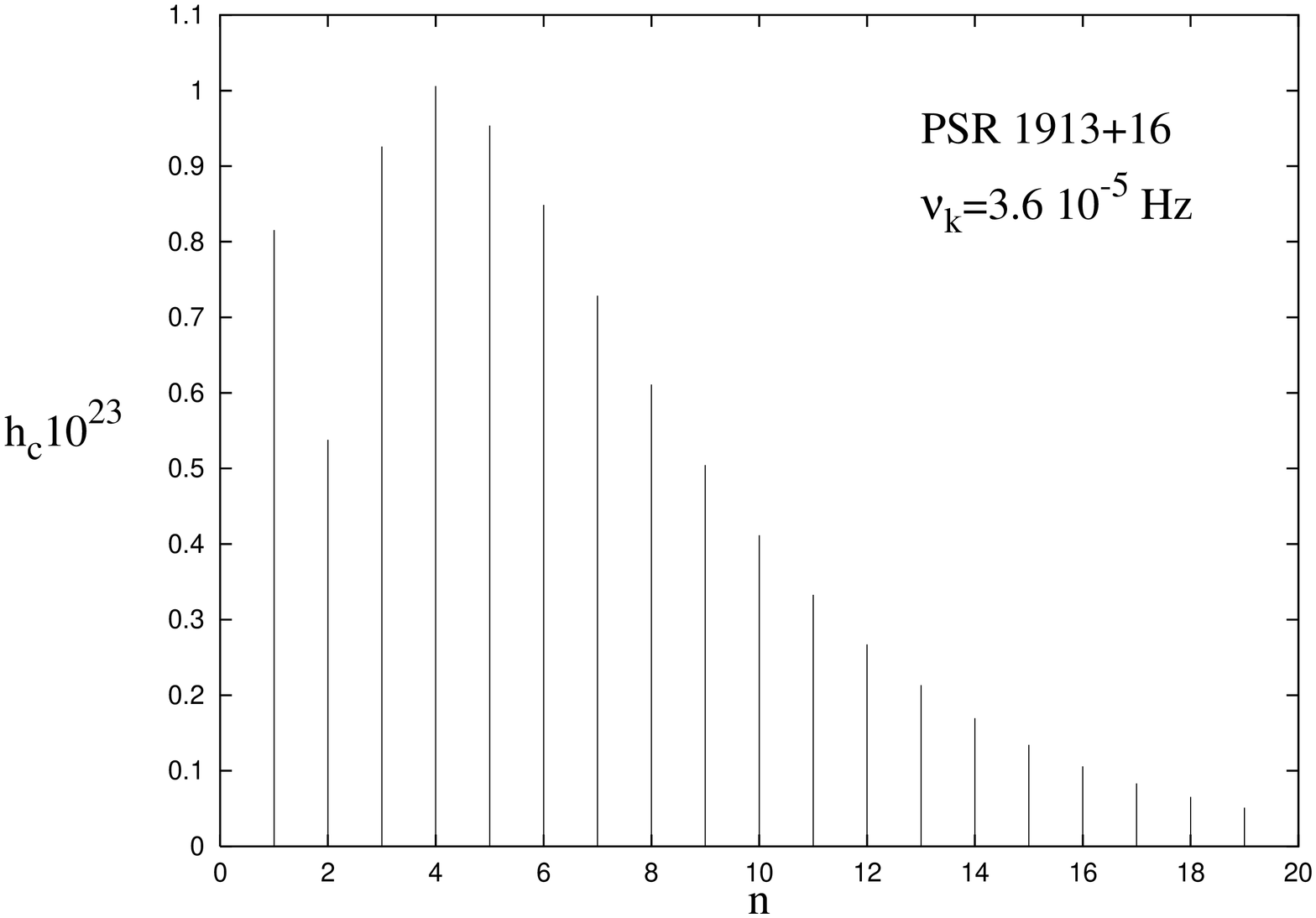,angle=270,width=9cm}}}
\vskip 10pt
\caption{
The characteristic amplitude computed from eqs. (\ref{hc})
is plotted versus the harmonic index $n$
for four selected EPS's and for the binary system PSR 1913+16. 
For the two systems in the upper panel,  
HD 283750 and $\tau$ Boo,
the planet moves on a nearly circular orbit, with
$e=0.02$ and $e=0.018$, respectively.
In this case the emission is concentrated at twice the 
keplerian frequency, and the
amplitude is comparable to the maximum wave amplitude emitted 
by the binary pulsar PSR 1913+16.
In the lower part of the figure 
\op h_c\cl is plotted
for two systems with high eccentricity: 
$16$ Cyg B ($e=0.634$) and HD $89707$   ($e=0.93$),
and for PSR 1913+16 ($e=0.617$)
In these cases the spectral lines are emitted  at frequencies 
multiple of \op\nu_k,$ and their number  increases with
$e$.
}
\label{fig2}
\end{figure}
The average energy flux $F_n$  relative to the $n-$th harmonic is 
 given by:
\be
F_n=\frac{c^3(n\omega_k)^2}{8\pi G}
\left[
\langle \tilde{h}^{(n)}_{\theta \theta}\rangle^2+
\langle \tilde{h}^{(n)}_{\theta \phi}\rangle^2
\right]
\ee
where $\langle \tilde{h}^{(n)}_{\theta \theta}\rangle^2$ and
$\langle \tilde{h}^{(n)}_{\theta \phi}\rangle^2$,
are the square of the $n-$th Fourier component of the two independent 
polarizations, averaged over the solid angle
$$\langle \tilde{h}^{(n)}_{\theta \theta}\rangle^2
=\frac{1}{4\pi}\int{d\Omega~\left|h_{\theta \theta}(n\omega_k,r,\theta,\phi)
\right|^2},\qquad
\langle \tilde{h}^{(n)}_{\theta \phi}\rangle^2=\frac{1}{4\pi}
\int{d\Omega~\left|h_{\theta \phi}(n\omega_k,r,\theta,\phi)\right|^2}.$$

An estimate of the characteristic amplitude of the gravitational waves 
emitted by a planetary system can now be  given by using
the well known formula\cite{Kip}
\be
\label{hc}
h_c(n \omega_k,r)=\sqrt{\frac{2}{3}}\left[
\langle \tilde{h}^{(n)}_{\theta \theta}\rangle^2+
\langle \tilde{h}^{(n)}_{\theta \phi}\rangle^2\right]^{1/2},
\ee
where the factor $\sqrt{2/3}$ takes into account the average over orientation.
The aforementioned procedure,
which is equivalent to that introduced by 
Peters and Mathews\cite{PetersMathews}, 
has been applied to compute the characteristic
wave amplitude \op h_c\cl   impinging on Earth, emitted by
our set of  EPS's and, for comparison, by the binary system PSR 1913+16.
Some results are shown in  figure 1, where \op h_c\cl is plotted
as a function of the harmonic index $n$.
It should be reminded that  PSR 1913+16 is composed of
two very compact stars with masses \op m_1=1.4411~\msun\cl
and \op m_2=1.3874~\msun,\cl revolving around  their center of mass
with an eccentric orbit ($e=0.617139$),  semimajor axis \op a=
1.9490\cdot 10^{12}\cl cm, and keplerian frequency
\op\nu_k=3.583\cdot10^{-5}~Hz.\cl
The binary system is at a distance  \op D=5\cl kpc from Earth.
In the upper panel of figure 1, we show two systems  
in which the planet moves in a nearly circular orbit around the central star,
HD 283750 ($e=0.02$) and $\tau$ Boo ($e=0.018$).
\begin{table}
\label{hcar}
\centering
\caption{The maximum characteristic amplitude of the waves emitted by the 
selected set of EPS's and by the binary system PSR 1913+16,
 due to their orbital motion.}
\begin{tabular}{@{}llll@{}}
\\
\multicolumn{1}{c} {Star} &$\nu_k~(Hz)$   &$h_{c~max}$ 
        &$\nu_{max}~(Hz)$  \\
\hline
HD 75289    	&$3.3\cdot10^{-6}$   &$2.1\cdot10^{-25}$  &$6.6\cdot10^{-6}$  \\
51 Peg      	&$2.8\cdot10^{-6}$   &$4.0\cdot10^{-25}$  &$5.5\cdot10^{-6}$  \\
$\upsilon$ And  &$2.5\cdot10^{-6}$  &$7.9\cdot10^{-26}$  &$5.0\cdot10^{-6}$ \\
55 Cnc      	&$7.9\cdot10^{-7}$    &$3.8\cdot10^{-25}$     &$1.6\cdot10^{-6}$ \\
$\rho$ CrB  	&$2.9\cdot10^{-7}$   &$1.5\cdot10^{-25}$  &$5.8\cdot10^{-7}$ \\
HD 210277   	&$2.6\cdot10^{-8}$   &$1.9\cdot10^{-26}$  &$7.9\cdot10^{-8}$ \\
16 Cyg B	&$1.4\cdot10^{-8}$   &$1.6\cdot10^{-26}$  &$5.8\cdot10^{-8}$ \\
Gl 876		&$1.9\cdot10^{-7}$   &$3.5\cdot10^{-25}$  &$3.8\cdot10^{-7}$ \\
47 Uma		&$1.1\cdot10^{-8}$   &$5.3\cdot10^{-26}$  &$2.1\cdot10^{-8}$ \\
14 Her		&$7.1\cdot10^{-9}$   &$2.7\cdot10^{-26}$  &$1.4\cdot10^{-8}$ \\
Gl 86		&$7.3\cdot10^{-7}$   &$1.6\cdot10^{-24}$  &$1.5\cdot10^{-6}$ \\
$\tau$ Boo	&$3.5\cdot10^{-6}$   &$6.6\cdot10^{-24}$  &$7.0\cdot10^{-6}$ \\
HD 168443	&$2.0\cdot10^{-7}$   &$1.6\cdot10^{-25}$  &$6.0\cdot10^{-7}$ \\
70 Vir		&$9.9\cdot10^{-8}$   &$3.6\cdot10^{-25}$  &$2.0\cdot10^{-7}$ \\
HD 114762	&$1.4\cdot10^{-7}$   &$2.0\cdot10^{-25}$  &$2.7\cdot10^{-7}$ \\
HD 110833	&$4.3\cdot10^{-8}$   &$2.6\cdot10^{-25}$  &$2.1\cdot10^{-7}$ \\
HD 112758	&$1.1\cdot10^{-7}$   &$3.0\cdot10^{-24}$  &$2.2\cdot10^{-7}$ \\
HD 29587	&$1.0\cdot10^{-8}$   &$2.3\cdot10^{-25}$  &$2.0\cdot10^{-8}$ \\
HD 283750	&$6.5\cdot10^{-6}$   &$3.7\cdot10^{-23}$  &$1.3\cdot10^{-5}$ \\
HD 89707	&$3.9\cdot10^{-8}$   &$8.4\cdot10^{-25}$  &$3.9\cdot10^{-8}$ \\
HD 217580       &$2.5\cdot10^{-8}$   &$8.5\cdot10^{-25}$  &$7.6\cdot10^{-8}$ \\
\hline 
\hline 
PSR 1913+16    &$3.6\cdot10^{-5}$  &$1.0\cdot10^{-23}$  &$1.4\cdot10^{-4}$ \\
\hline 
\hline 
\end{tabular}
\end{table}
As expected, the emission is concentrated at twice the keplerian 
frequency, and the amplitude is comparable to the maximum wave 
amplitude reached by  PSR 1913+16, which is shown for
comparison at the bottom of the figure. However, the frequency
is about ten times lower than that of the binary pulsar.
In the lower panel  we show the characteristic emission 
of two EPS's with high eccentricity, $16$ Cyg B 
($e=0.634$) and HD $89707$   ($e=0.93$).
In this case the gravitational emission  is spread over a larger set of
frequencies, all multiple of \op\nu_k,$ as for PSR 1913+16.

In table 3 we tabulate
the keplerian frequency, the maximum characteristic amplitude on Earth,
and the corresponding frequency, for each EPS. 
In the last row  the same data  are listed
for PSR 1913+16.
For most systems the maximum emission frequency appears to be extremely
low, in general smaller than \op 10^{-6}~Hz\cl except  the case of
HD 283750, for which \op \nu_{max}=1.3\cdot 10^{-5}~Hz.\cl

\section{Resonant excitation of the modes of a star by the orbiting
planet}

We shall now consider another mechanism through which gravitational waves
can be emitted by a system composed of a star and a planet.
Since the masses of planets are much smaller than those of stars,
in what follows we shall treat the planet as a pointlike mass,
whereas the central star  will be assumed
to have a structure, and to possess a set of eigenmodes of
oscillation, the quasi-normal modes, that are associated to
the emission of  gravitational radiation.  
We want  to establish  whether these modes can be excited  by a planet. 

As mentioned  in the introduction, perturbative calculations in the
framework of general relativity made by Kojima\cite{kojima}
have shown that,  when a small mass moves in a circular orbit 
with  orbital frequency \op \omega_k\cl around a compact star,
the fundamental mode is excited if its frequency is
\op \omega_f = 2\omega_k.\cl
In this case, a sharp resonance occurs,
and the characteristic wave amplitude can be
considerably larger than that evaluated by the quadrupole formula.
It may be reminded that the quadrupole
radiation by a planet in circular orbit around a star
is also emitted at twice the orbital frequency, thus 
the condition of resonant excitation can also be formulated in terms 
of the  quadrupole emission frequencies.

Similar calculations have never been done for solar type stars;
however, Kojima's results  suggest 
that the  resonant excitation of a mode 
may enhance the gravitational emission also for non compact stars.  
For this reason, it  is interesting to check whether
the conditions of resonant  excitation
can be fulfilled in the planetary systems listed in table 1.  
We shall first verify 
whether the maximum quadrupole emission frequency  of the planets
of our set of EPS's, \op\nu_{max}\cl (table 3, column 4) is  close enough to
any of the frequencies of the  modes of the central star.

The oscillation frequencies  of
a star can be computed if we  make an assumption on its internal
structure, i.e. on the equation of state prevailing in the interior. 
We shall consider, as an example, a very simple, polytropic model of
star, with polytropic index \op n=2.\cl
The oscillation frequencies of newtonian polytropic stars 
are known to  scale with the mean density of the star\cite{cox}.
In table 4
we tabulate the dimensionless eigenfrequencies  of the modes,
\op \nu_{mode}/(GM_\star/R_\star^{3})^{1/2},\cl for the  chosen value of $n$.
In order to explicitely compute the frequency of a given mode for 
a given  star, the entries of table 4
have to be multiplied by the entries of table 1, column 5.
For instance, for the star HD 89707 
the frequency of the $f$-mode is given by
$\nu=0.28\times 6.1\cdot 10^{-4}=
1.7\cdot 10^{-4}~Hz.\cl
This number has to be compared whith 
\op \nu_{max}=2.2\cdot 10^{-6}~Hz,\cl
given in table 3 for the same star, which is much
smaller. This means that the planet cannot excite the $f$-mode of the
central star.
\begin{table}
\centering
\caption{The dimensionless eigenfrequencies of the modes of a polytropic
stars are tabulated for the polytropic index $n=2$.}
\begin{tabular}{@{}lllllll@{}}
\\
\hline
\multicolumn{1}{c}{mode}      &$\nu_{mode}/(GM_\star/R_\star^{3})^{1/2}$   
    &&{mode}        &$\nu_{mode}/(GM_\star/R_\star^{3})^{1/2}$          \\
\hline
&$p_{10}$         &2.58        &&$f$              &0.28           \\
&$p_{9}$          &2.37        &&$g_{1}$          &0.119           \\
&$p_{8}$          &2.15        &&$g_{2}$          &0.087           \\
&$p_{7}$          &1.93        &&$g_{3}$          &0.068            \\
&$p_{6}$          &1.70        &&$g_{4}$          &0.056            \\
&$p_{5}$          &1.47        &&$g_{5}$          &0.048           \\
&$p_{4}$          &1.24        &&$g_{6}$          &0.042           \\
&$p_{3}$          &1.01        &&$g_{7}$          &0.037           \\
&$p_{2}$          &0.78        &&$g_{8}$          &0.033           \\
&$p_{1}$          &0.54        &&$g_{9}$          &0.030           \\
&                 &            &&$g_{10}$         &0.028            \\
\hline
\end{tabular}
\end{table}
By repeating the same calculation for all systems and all modes,
we find that it is unlikely that the stellar 
modes are excited in the EPS's we consider.
This is because the angular
velocities reached by the planets are too low, and
consequently the frequencies of the radiation they emit
are  lower than those of the $f$-mode or of the lowest-order
$g$-modes of the star.
Higher order $g$-modes could be excited, but the 
efficiency in producing gravitational radiation by this process
would be too low.

However, it should be reminded that according to  recent
theories on the evolution of planetary systems\cite{Ford},  
there could exist planets moving on orbits even closer to the 
central star than those observed until now.
In view of this possibility, it is interesting to investigate 
whether it is  possible 
for a planet to  approach a star at such a short distance that its  
angular velocity is high enough to
excite the $f$-mode or the  lowest-order $g$-modes,  without being
disrupted by  tidal forces.  In addition, we shall impose
that the star does not accrete matter onto the planet. These conditions are
equivalent to impose that neither the planet nor the star
overflow their Roche lobe.
We would like to stress that this limit takes into account only the
gravitational interaction between the planet and the star. There may exist
other processes that would prevent the planet to reach the innermost orbit
allowed by the Roche-lobe analysis. However, as far as we know, the
present knowledge on the formation of planets and on their possible migration
toward the central star, still does not allow to firmly establish what is
the minimum distance from a star at which a planet can safely sit. 

We shall assume, for simplicity, that the planet flies on a circular orbit
with keplerian angular velocity 
\op\omega_k,\cl
and consequently emits radiation at the frequency
\op \omega^{GW}=2 \omega_k.\cl
Let us indicate the dimensionless frequency tabulated in table 4
multiplied by \op 2\pi\cl as \op k_{mode}:\cl
for instance, \op k_{g_{1}}=0.750.\cl
The condition that a mode of the star is excited by the  resonant
interaction with the planet, 
\op \omega_{mode}= 2 \omega_k,\cl
 therefore becomes
\be
k_{mode}\cdot\left[\frac{G M_\star}{R_\star^3}\right]^{1/2}=
2 \left[\frac{G\left(M_p+M_\star\right)}{a^3}\right]^{1/2},
\label{cond1}
\ee
which  can be written as
\be
a=\left[
\frac{4}{k^2_{mode}}\cdot
\left(1+\frac{M_p}{M_\star}\right)
\right]^{1/3}R_\star.
\label{cond2}
\ee
This equation gives
the value the  separation star-planet must have  for a given
mode to be excited.

The further condition  that the planet  lies inside
its Roche lobe,  \op R_p < R_{RL},\cl  is equivalent to the following
constraint on  its density: 
\[ 
\rho_p > \rho_{RL},\]
where \op\dps{ \rho_{RL} = \frac{M_p}{\frac{4}{3}\pi R_{RL}^3}}\cl
is the critical density.  If we now  introduce the dimensionless quantity 
\op \overline{R}_{RL}\cl as given by
\be
R_{RL}=a~\overline{R}_{RL},
\label{rbar}
\ee
i.e. we set to 1 the radius of the orbit, by the use of 
eq. (\ref{cond2}) we find 
\be
\rho_{RL}=\frac{M_p}{
\frac{4}{3}\pi R_\star^3
\left[\dps{\frac{4}{k^2_{mode}}\cdot
\left(1+\frac{M_p}{M_\star}\right)}
\right]\overline{R}_{RL}^3},
\ee
which can be rewritten as
\be
\frac{\rho_{RL}}{\rho_\star}
=k^2_{mode}\cdot \frac{M_p/M_\star}{4\left(1+M_p/M_\star\right)
\overline{R}_{RL}^3}
\label{cond4}
\ee
where \op\rho_\star\cl is the mean density of the central star.
Thus, a planet can  excite a mode of the star corresponding to an assigned 
\op k_{mode},\cl  without overflowing
its Roche lobe, if the ratio between its mean density  and that
of the central star  exceeds the  critical ratio (\ref{cond4}).

We have computed the dimensionless
radius of the Roche lobe \op \overline{R}_{RL}\cl
and the corresponding critical ratio (\ref{cond4}),
for assigned values of the ratio \op M_p/M_\star,\cl
and the results  are shown in table \ref{exci}, 
which has to be read as follows.
\begin{table}
\centering
\caption{Ratios of the critical density of the planet $\rho_{RL}$ 
to the central star density $\rho_*$ for different values of the mass 
ratio, $M_{p}/M_{\star}$, and different oscillation modes of a 
polytropic star with $n=2$ (see text).}
\begin{tabular}{@{}llllllll@{}}
\\
                         &\multicolumn{6}{c}{$M_{p}/M_{\star}$} \\
\hline
  &$10^{-6}$  &$10^{-5}$  &$10^{-4}$  &$10^{-3}$  &$10^{-2}$  &$10^{-1}$ \\
\hline
$f$        &7.89    &7.95       &8.04       &-          &-          &-    \\
$g_{1}$    &1.43    &1.44       &1.45       &1.49       &1.58       &1.83 \\
$g_{2}$    &0.75    &0.76       &0.77       &0.79       &0.83       &0.96 \\
$g_{3}$    &0.47    &0.47       &0.47       &0.49       &0.52       &0.60 \\
$g_{4}$    &0.32    &0.32       &0.32       &0.33       &0.35       &0.41\\
$g_{5}$    &0.23    &0.23       &0.23       &0.24       &0.26       &0.30 \\
$g_{6}$    &0.17    &0.18       &0.18       &0.18       &0.19       &0.22 \\
$g_{7}$    &0.14    &0.14       &0.14       &0.14       &0.15       &0.18 \\
$g_{8}$    &0.11    &0.11       &0.11       &0.12       &0.12       &0.14 \\
$g_{9}$    &0.09    &0.09       &0.09       &0.10       &0.10       &0.12 \\
$g_{10}$   &0.08    &0.08       &0.08       &0.08       &0.09       &0.10 \\
\hline
\end{tabular}
\label{exci}
\end{table}  
Suppose that the ratio between the mass of a planet and that of
the central star is \op M_p/M_\star=10^{-3}.\cl 
The frequency of the 
quadrupole radiation emitted by the planet will coincide with that 
of the first $g$-mode of the star, and the planet will not be disrupted,
only if its density is  higher that \op 1.49~\rho_\star.\cl
We have also checked whether the star  overflows its Roche lobe and
accretes matter onto the planet, and excluded from table \ref{exci}
the corresponding cases.
It should be reminded that
the ratio between the mean density of the planets of the solar
system and that of the  Sun  is  3.9 for Mercury and the Earth, 3.7 for Venus,
2.8 for Mars, 0.9 for Jupiter, etc.  A comparison of these values 
with the  data of table \ref{exci} suggests
that in principle, there can exist EPS's in which the first 
$g$-modes could be excited by a resonant process. 
For instance, a planet like the Earth could approach
a polytropic  star ($n=2$)
with the mass of  the sun at a distance close 
enough to excite the mode $g_1$
without being disrupted by the tidal interaction, 
whereas a planet like Jupiter could  only be at a distance 
good to excite the second $g$-mode.

By the Roche-lobe analysis we can also deduce  another interesting
information.
Suppose that a planetary system made of a star with the mass
of the Sun and a planet in circular orbit, 
is located at a fiducial distance \op D=10~pc.\cl
We do not make any assumption on the internal structure of the star, but
assign the values of the mass and of the mean density
of the planet.  In particular we consider four 
planets, with  mass and density equal to that of Mercury, 
of the Earth, of Jupiter and one with a mass equal to 13 times the mass of
Jupiter and similar mean density\cite{SHBG}.
We want to answer the following questions:
\begin{itemize}
\item{} what is the minimum radius of the orbit?
\item{}what is the corresponding quadrupole emission 
frequency \op\nu^{GW}=2\nu_k?\cl
\item{}what is the corresponding characteristic amplitude on Earth?
\end{itemize}
The answer is in table 6, where the required data are given for
the  four planets.

\begin{table}
\centering
\caption{We tabulate the minimum radius of the circular orbit, $a_{min}$,
the quadrupole emission frequency, $\nu^{GW}$, and the characteristic
amplitude of the corresponding wave emitted by four planets orbiting
around a star  with the mass of the Sun. We assume that
the planets have mass and density equal to those of Mercury,  
Earth and Jupiter. The last planet has  13 times the mass  of Jupiter
and the same density.  These data
are obtained by imposing that the planets lie inside their
Roche lobe (see text), and that the system is located at $D=10~pc$ from
Earth.}
\begin{tabular}{@{}lllll@{}}
\\
\hline
Type of planet           &     &$a_{min}$ (cm)      &$\nu^{GW}=2\nu_k$  &$h_c$  \\
\hline
Mercury    &    &$9.6\cdot 10^{10}$ &$1.2\cdot10^{-4}$&$1.8\cdot10^{-27}$\\
Earth      &    &$9.6\cdot 10^{10}$ &$1.2\cdot10^{-4}$&$3.2\cdot10^{-26}$\\
Jupiter    &    &$1.6\cdot 10^{11}$ &$5.9\cdot10^{-5}$&$6.2\cdot10^{-24}$\\
13$\times$Jupiter&   &$1.6\cdot 10^{11}$ &$5.7\cdot10^{-5}$&$7.9\cdot10^{-23}$\\
\hline
\end{tabular}
\label{fin}
\end{table}  
From table 6  we see that planets like Jupiter or bigger could
emit quadrupole radiation at a frequency in the  bandwidth of space 
interferometers, and with an amplitude which could be even ten times bigger
than that emitted by the binary pulsar PSR 1913+16. In addition, if
the quadrupole radiation  resonates with a mode  of the star, 
the amount of emitted energy could be even larger.

\section{Concluding Remarks}
In this paper we have studied some of the EPS's discovered up to now,
with the aim of characterizing their gravitational emission.

As far as the quadrupole emission is concerned, we have shown that 
among those systems there is one, HD 283750, that emits a signal 
whith a maximum amplitude on Earth even higher than that of the binary pulsar 
PSR 1913+16, but at a frequency which is six times smaller. 
Moreover, since the orbit of the planet is nearly circular, the radiation 
is almost entirely emitted at twice the keplerian frequency, whereas 
for systems with high eccentricity, as the binary pulsar, the radiation 
is emitted also at higher multiples of \op \omega_k.\cl

A further mechanism of gravitational emission  is
the resonant excitation of a mode of the star due to the tidal interaction
with a planet. 
For compact stars, this mechanism has proved to be much more efficient
than the quadrupole emission, thus it is interesting to check
whether the conditions needed for the
resonant excitation of the $f$- and $g$-modes 
can be fulfilled in any of the observed EPS's.
Although there exist planets that have a very short orbital period 
(up to 1.79 days), and orbital frequencies  close to
\op  10^{-5}~Hz,\cl these values are 
too low to induce a  resonant excitation of the  $f$-mode or  the lowest 
$g$-modes  of a solar type star.

However, since many new EPS's will certainly be discovered in the 
near future, we wanted to understand whether
more favourable conditions 
for the emission of gravitational waves may occur in these systems. 
In particular, since a higher emission frequency would be desirabile 
both for the excitation of the lowest order $g$-modes
of the central star, and for a possible detection by space 
interferometers, we have investigated how close can a planet  move
on a circular orbit around a star without being tidally disrupted, 
and without accreting matter from the star.

It should be stressed that the limits we establish by the Roche-lobe
analysis do not take into account other  processes which may
destabilize the orbit of the planets, whose study is beyond the scopes of
this paper. Having this caveat in mind,
we have established that, in principle, there could exist systems 
in which the excitability conditions of the lowest order $g$-modes 
could  be fulfilled. 
In this case radiation would be emitted at frequencies of the order of
\op \sim 10^{-4}~Hz.\cl 
The amplitude of the emitted gravitational signals depends on the ratio 
\op M_p/M_\star,\cl  and
for a planet  like Jupiter or bigger, located at a distance of
$10~pc$, it would range between \op 10^{-23}-
10^{-22} .\cl 
The emitted radiation could be even larger, if the system is in a
condition of resonant excitation of a mode of the star, and
we plan  to investigate this problem in detail in a subsequent paper.

\vskip 10pt

{\bf{Acknowledgements}}

\vskip 5pt

We would like to thank  K. Kokkotas and  L. Stella for 
their valuable suggestions.

\nonumsection{References}

\end{document}